\font\cmss=cmss10 
\font\cmsss=cmss10 at 7pt

\font\bigit=cmti10 scaled \magstep1

\def\unlockat{\catcode`\@=11}

\unlockat

\global\newcount\secno \global\secno=0
\global\newcount\prono \global\prono=0
\def\newsec#1{\global\advance\secno by1\message{(\the\secno. #1)}
\global\subsecno=0\global\subsubsecno=0
\global\deno=0\global\teno=0
\eqnres@t\noindent
{\titlefont\the\secno. #1}
\writetoca{{\bf\secsym} { #1}}\par\nobreak\medskip\nobreak}
\global\newcount\subsecno \global\subsecno=0
\def\subsec#1{\global\advance\subsecno
by1\message{(\secsym\the\subsecno. #1)}
\ifnum\lastpenalty>9000\else\bigbreak\fi
\global\subsubsecno=0
\global\deno=0
\global\teno=0
\noindent{\bf\secsym\the\subsecno. #1}
\writetoca{\bf \string\quad {\secsym\the\subsecno.} {\it  #1}}
\par\nobreak\medskip\nobreak}
\global\newcount\subsubsecno \global\subsubsecno=0
\def\subsubsec#1{\global\advance\subsubsecno by1
\message{(\secsym\the\subsecno.\the\subsubsecno. #1)}
\ifnum\lastpenalty>9000\else\bigbreak\fi
\noindent\quad{\bf \secsym\the\subsecno.\the\subsubsecno.}{\ \sl \ #1}
\writetoca{\string\qquad\bf { \secsym\the\subsecno.\the\subsubsecno.}{\sl  \ #1}}
\par\nobreak\medskip\nobreak}

\global\newcount\deno \global\deno=0
\def\de#1{\global\advance\deno by1
\message{(\bf Definition\quad\secsym\the\subsecno.\the\deno #1)}
\ifnum\lastpenalty>9000\else\bigbreak\fi
\noindent{\bf Definition\quad\secsym\the\subsecno.\the\deno}{#1}
\writetoca{\string\qquad{\secsym\the\subsecno.\the\deno}{#1}}}

\global\newcount\prono \global\prono=0
\def\pro#1{\global\advance\prono by1
\message{(\bf Proposition\quad\secsym\the\subsecno.\the\prono 
)}
\ifnum\lastpenalty>9000\else\bigbreak\fi
\noindent{\bf Proposition\quad
\the\prono\quad}{\ninepoint #1}
}

\global\newcount\teno \global\prono=0
\def\te#1{\global\advance\teno by1
\message{(\bf Theorem\quad\secsym\the\subsecno.\the\teno #1)}
\ifnum\lastpenalty>9000\else\bigbreak\fi
\noindent{\bf Theorem\quad\secsym\the\subsecno.\the\teno}{#1}
\writetoca{\string\qquad{\secsym\the\subsecno.\the\teno}{#1}}}
\def\subsubseclab#1{\DefWarn#1\xdef #1{\noexpand\hyperref{}{subsubsection}%
{\secsym\the\subsecno.\the\subsubsecno}%
{\secsym\the\subsecno.\the\subsubsecno}}%
\writedef{#1\leftbracket#1}\wrlabeL{#1=#1}}

\def\unredoffs{} \def\redoffs{\voffset=-.40truein\hoffset=-.40truein}
\def\speclscape{}

\newbox\leftpage \newdimen\fullhsize \newdimen\hstitle \newdimen\hsbody
\tolerance=1000\hfuzz=2pt

\catcode`\@=11
\def\bigans{b }
\def\answ{b }

\ifx\answ\bigans\message{(This will come out unreduced.}
\magnification=1200\unredoffs\baselineskip=16pt plus 2pt minus 1pt
\hsbody=\hsize \hstitle=\hsize

\else\message{(This will be reduced.} \let\l@r=L
\magnification=1200\baselineskip=16pt plus 2pt minus 1pt \vsize=7truein
\redoffs \hstitle=8truein\hsbody=4.75truein\fullhsize=10truein\hsize=\hsbody
\output={\ifnum\pageno=0

   \shipout\vbox{{\hsize\fullhsize\makeheadline}
     \hbox to \fullhsize{\hfill\pagebody\hfill}}\advancepageno
   \else
   \almostshipout{\leftline{\vbox{\pagebody\makefootline}}}\advancepageno
   \fi}
\def\almostshipout#1{\if L\l@r \count1=1 \message{[\the\count0.\the\count1]}
       \global\setbox\leftpage=#1 \global\let\l@r=R
  \else \count1=2
   \shipout\vbox{\speclscape{\hsize\fullhsize\makeheadline}
       \hbox to\fullhsize{\box\leftpage\hfil#1}}  \global\let\l@r=L\fi}
\fi

\def\Title#1#2{
\abstractfont\hsize=\hstitle\rightline{#1}%
\vskip 5pt\centerline{\titlefont #2}\abstractfont\vskip .5in\pageno=0}
%


\def\draftmode{\message{ DRAFTMODE }\def\draftdate{{\rm preliminary draft:
\number\month/\number\day/\number\yearltd\ \ \hourmin}}%

\writelabels\baselineskip=20pt plus 2pt minus 2pt
  {\count255=\time\divide\count255 by 60 \xdef\hourmin{\number\count255}
   \multiply\count255 by-60\advance\count255 by\time
   \xdef\hourmin{\hourmin:\ifnum\count255<10 0\fi\the\count255}}}

\def\nolabels{\def\wrlabeL##1{}\def\eqlabeL##1{}\def\reflabeL##1{}}
\def\writelabels{\def\wrlabeL##1{\leavevmode\vadjust{\rlap{\smash%
{\line{{\escapechar=` \hfill\rlap{\sevenrm\hskip.03in\string##1}}}}}}}%
\def\eqlabeL##1{{\escapechar-1\rlap{\sevenrm\hskip.05in\string##1}}}%
\def\reflabeL##1{\noexpand\llap{\noexpand\sevenrm\string\string\string##1}}}
\nolabels
%


\global\newcount\secno \global\secno=0
\global\newcount\meqno
\global\meqno=1
\def\eqnres@t{\xdef\secsym{\the\secno.}\global\meqno=1
\bigbreak\bigskip}
\def\sequentialequations{\def\eqnres@t{\bigbreak}}
\def\appendix#1#2{\vfill\eject\global\meqno=1\global\subsecno=0\xdef\secsym{\hbox{#1.}}
\bigbreak\bigskip\noindent{\bf Appendix #1. #2}\message{(#1. #2)}
\writetoca{Appendix {#1.} {#2}}\par\nobreak\medskip\nobreak}

\def\eqnn#1{\xdef #1{(\secsym\the\meqno)}\writedef{#1\leftbracket#1}%
\global\advance\meqno by1\wrlabeL#1}
\def\eqna#1{\xdef #1##1{\hbox{$(\secsym\the\meqno##1)$}}
\writedef{#1\numbersign1\leftbracket#1{\numbersign1}}%
\global\advance\meqno by1\wrlabeL{#1$\{\}$}}
\def\eqn#1#2{\xdef #1{(\secsym\the\meqno)}\writedef{#1\leftbracket#1}%
\global\advance\meqno by1$$#2\eqno#1\eqlabeL#1$$}

\newskip\footskip\footskip14pt plus 1pt minus 1pt

\def\footnotefont{\ninepoint}\def\f@t#1{\footnotefont #1\@foot}
\def\f@@t{\baselineskip\footskip\bgroup\footnotefont\aftergroup\@foot\let\next}
\setbox\strutbox=\hbox{\vrule height9.5pt depth4.5pt width0pt}
\global\newcount\ftno \global\ftno=0
\def\foot{\global\advance\ftno by1\footnote{$^{\the\ftno}$}}

\newwrite\ftfile
\def\footend{\def\foot{\global\advance\ftno by1\chardef\wfile=\ftfile
$^{\the\ftno}$\ifnum\ftno=1\immediate\openout\ftfile=foots.tmp\fi%
\immediate\write\ftfile{\noexpand\smallskip%
\noexpand\item{f\the\ftno:\ }\pctsign}\findarg}%
\def\footatend{\vfill\eject\immediate\closeout\ftfile{\parindent=20pt
\centerline{\bf Footnotes}\nobreak\bigskip\input foots.tmp }}}
\def\footatend{}

\global\newcount\refno \global\refno=1
\newwrite\rfile
\def\ref{[\the\refno]\nref}
\def\nref#1{\xdef#1{[\the\refno]}\writedef{#1\leftbracket#1}%
\ifnum\refno=1\immediate\openout\rfile=refs.tmp\fi \global\advance\refno
by1\chardef\wfile=\rfile\immediate \write\rfile{\noexpand\item{#1\
}\reflabeL{#1\hskip.31in}\pctsign}\findarg}

\def\findarg#1#{\begingroup\obeylines\newlinechar=`\^^M\pass@rg}
{\obeylines\gdef\pass@rg#1{\writ@line\relax #1^^M\hbox{}^^M}%
\gdef\writ@line#1^^M{\expandafter\toks0\expandafter{\striprel@x #1}%
\edef\next{\the\toks0}\ifx\next\em@rk\let\next=\endgroup\else\ifx\next\empty%
\else\immediate\write\wfile{\the\toks0}\fi\let\next=\writ@line\fi\next\relax}}
\def\striprel@x#1{} \def\em@rk{\hbox{}}
\def\lref{\begingroup\obeylines\lr@f}
\def\lr@f#1#2{\gdef#1{\ref#1{#2}}\endgroup\unskip}
\def\semi{;\hfil\break}
\def\addref#1{\immediate\write\rfile{\noexpand\item{}#1}}

\def\startrefs#1{\immediate\openout\rfile=refs.tmp\refno=#1}
\def\xref{\expandafter\xr@f}\def\xr@f[#1]{#1}
\def\refs#1{\count255=1[\r@fs #1{\hbox{}}]}
\def\r@fs#1{\ifx\und@fined#1\message{reflabel \string#1 is undefined.}%
\nref#1{need to supply reference \string#1.}\fi%
\vphantom{\hphantom{#1}}\edef\next{#1}\ifx\next\em@rk\def\next{}%
\else\ifx\next#1\ifodd\count255\relax\xref#1\count255=0\fi%
\else#1\count255=1\fi\let\next=\r@fs\fi\next}


\def\writetoc{\immediate\openout\tfile=BAtoc.tmp
    \def\writetoca##1{{\edef\next{\write\tfile{\noindent  ##1
    \string\leaderfill {\noexpand\number\pageno} \par}}\next}}}


%
\catcode`\@=12 
%
\edef\tfontsize{\ifx\answ\bigans scaled\magstep3\else scaled\magstep4\fi}
\font\titlerm=cmr10 \tfontsize \font\titlerms=cmr7 \tfontsize
\font\titlermss=cmr5 \tfontsize \font\titlei=cmmi10 \tfontsize
\font\titleis=cmmi7 \tfontsize \font\titleiss=cmmi5 \tfontsize
\font\titlesy=cmsy10 \tfontsize \font\titlesys=cmsy7 \tfontsize
\font\titlesyss=cmsy5 \tfontsize \font\titleit=cmti10 \tfontsize
\skewchar\titlei='177 \skewchar\titleis='177 \skewchar\titleiss='177
\skewchar\titlesy='60 \skewchar\titlesys='60 \skewchar\titlesyss='60
\def\titlefont{\def\rm{\fam0\titlerm}
\textfont0=\titlerm \scriptfont0=\titlerms \scriptscriptfont0=\titlermss
\textfont1=\titlei \scriptfont1=\titleis \scriptscriptfont1=\titleiss
\textfont2=\titlesy \scriptfont2=\titlesys \scriptscriptfont2=\titlesyss
\textfont\itfam=\titleit
\def\it{\fam\itfam\titleit}\rm}
\font\authorfont=cmcsc10 \ifx\answ\bigans\else scaled\magstep1\fi
\ifx\answ\bigans\def\abstractfont{\tenpoint}\else \font\abssl=cmsl10 scaled
\magstep1 \font\absrm=cmr10 scaled\magstep1 \font\absrms=cmr7
scaled\magstep1 \font\absrmss=cmr5 scaled\magstep1 \font\absi=cmmi10
scaled\magstep1 \font\absis=cmmi7 scaled\magstep1 \font\absiss=cmmi5
scaled\magstep1 \font\abssy=cmsy10 scaled\magstep1 \font\abssys=cmsy7
scaled\magstep1 \font\abssyss=cmsy5 scaled\magstep1 \font\absbf=cmbx10
scaled\magstep1 \skewchar\absi='177 \skewchar\absis='177
\skewchar\absiss='177 \skewchar\abssy='60 \skewchar\abssys='60
\skewchar\abssyss='60
\def\abstractfont{\def\rm{\fam0\absrm}
\textfont0=\absrm \scriptfont0=\absrms \scriptscriptfont0=\absrmss
\textfont1=\absi \scriptfont1=\absis \scriptscriptfont1=\absiss
\textfont2=\abssy \scriptfont2=\abssys \scriptscriptfont2=\abssyss
\textfont\itfam=\bigit \def\it{\fam\itfam\bigit}\def\footnotefont{\tenpoint}%
\textfont\slfam=\abssl \def\sl{\fam\slfam\abssl}%
\textfont\bffam=\absbf \def\bf{\fam\bffam\absbf}\rm}\fi
\def\tenpoint{\def\rm{\fam0\tenrm}
\textfont0=\tenrm \scriptfont0=\sevenrm \scriptscriptfont0=\fiverm
\textfont1=\teni  \scriptfont1=\seveni  \scriptscriptfont1=\fivei
\textfont2=\tensy \scriptfont2=\sevensy \scriptscriptfont2=\fivesy
\textfont\itfam=\tenit \def\it{\fam\itfam\tenit}\def\footnotefont{\ninepoint}%
\textfont\bffam=\tenbf
\def\bf{\fam\bffam\tenbf}\def\sl{\fam\slfam\tensl}\rm}
\font\ninerm=cmr9 \font\sixrm=cmr6 \font\ninei=cmmi9 \font\sixi=cmmi6
\font\ninesy=cmsy9 \font\sixsy=cmsy6 \font\ninebf=cmbx9 \font\nineit=cmti9
\font\ninesl=cmsl9 \skewchar\ninei='177 \skewchar\sixi='177
\skewchar\ninesy='60 \skewchar\sixsy='60
\def\ninepoint{\def\rm{\fam0\ninerm}
\textfont0=\ninerm \scriptfont0=\sixrm \scriptscriptfont0=\fiverm
\textfont1=\ninei \scriptfont1=\sixi \scriptscriptfont1=\fivei
\textfont2=\ninesy \scriptfont2=\sixsy \scriptscriptfont2=\fivesy
\textfont\itfam=\ninei \def\it{\fam\itfam\nineit}\def\sl{\fam\slfam\ninesl}%
\textfont\bffam=\ninebf \def\bf{\fam\bffam\ninebf}\rm}
%
%

\hyphenation{anom-aly anom-alies coun-ter-term coun-ter-terms}
\def\inv{^{\raise.15ex\hbox{${\scriptscriptstyle -}$}\kern-.05em 1}}

\def\Dsl{\,\raise.15ex\hbox{/}\mkern-13.5mu D} 
\def\dsl{\raise.15ex\hbox{/}\kern-.57em\partial}

\def\tr#1{\, {\rm tr}\, \left( #1 \right)}

\def\lspace{\ifx\answ\bigans{}\else\qquad\fi}
\def\lbspace{\ifx\answ\bigans{}\else\hskip-.2in\fi} 
\def\boxeqn#1{\vcenter{\vbox{\hrule\hbox{\vrule\kern3pt\vbox{\kern3pt
     \hbox{${\displaystyle #1}$}\kern3pt}\kern3pt\vrule}
    }}}
\def\mbox#1#2{\vcenter{\hrule \hbox{\vrule height#2in
         \kern#1in \vrule} \hrule}}  


\newwrite\ffile\global\newcount\figno \global\figno=1
\def\nfig#1{\xdef#1{fig.~\the\figno}%
\writedef{#1\leftbracket fig.\noexpand~\the\figno}%
\ifnum\figno=1\immediate\openout\ffile=figs.tmp\fi\chardef\wfile=\ffile%
\immediate\write\ffile{\noexpand\medskip\noexpand\item{Fig.\ \the\figno. }
\reflabeL{#1\hskip.55in}\pctsign}\global\advance\figno by1\findarg}
\def\xfig{\expandafter\xf@g}
\def\xf@g fig.\penalty\@M\ {}
\def\figs#1{figs.~\f@gs #1{\hbox{}}}
\def\f@gs#1{\edef\next{#1}\ifx\next\em@rk\def\next{}\else
\ifx\next#1\xfig #1\else#1\fi\let\next=\f@gs\fi\next}
\newwrite\lfile
{\escapechar-1\xdef\pctsign{\string\%}\xdef\leftbracket{\string\{}
\xdef\rightbracket{\string\}}\xdef\numbersign{\string\#}}

\def\writestop{\def\writestoppt{\immediate\write\lfile{\string\pageno%
\the\pageno\string\startrefs\leftbracket\the\refno\rightbracket%
\string\def\string\secsym\leftbracket\secsym\rightbracket%
\string\secno\the\secno\string\meqno\the\meqno}\immediate\closeout\lfile}}
\def\writestoppt{}\def\writedef#1{}
\def\seclab#1{\xdef #1{\the\secno}\writedef{#1\leftbracket#1}\wrlabeL{#1=#1}}
\def\subseclab#1{\xdef #1{\secsym\the\subsecno}%
\writedef{#1\leftbracket#1}\wrlabeL{#1=#1}}
\newwrite\tfile \def\writetoca#1{}
\def\leaderfill{\leaders\hbox to 1em{\hss.\hss}\hfill}


\def\tilde{\widetilde}
\def\bar{\overline}
\def\hat{\widehat}
\def\ii{\rm i}

\def\cech{${\rm C}^{\kern-6pt \vbox{\hbox{$\scriptscriptstyle\vee$}\kern2.5pt}}${\rm ech}}
\def\Cech{${\sl C}^{\kern-6pt \vbox{\hbox{$\scriptscriptstyle\vee$}\kern2.5pt}}${\sl ech}}


\def\a{{\alpha}}

\def\e{{\epsilon}}

\def\ve{{\varepsilon}}

\def\l{{\lambda}}
\def\m{{\mu}}

\def\s{{\sigma}}
\def\t{{\theta}}


\def\p{\partial}

\def\inv{^{\raise.15ex\hbox{${\scriptscriptstyle -}$}\kern-.05em 1}}

\def\Dsl{\,\raise.15ex\hbox{/}\mkern-13.5mu D}
\def\dsl{\raise.15ex\hbox{/}\kern-.57em\partial}


\def\bA{{\bf A}}
\def\ba{{\bf a}}

\def\bg{{\bf g}}

\def\bi{{\bf i}}

\def\bK{{\bf K}}

\def\bM{{\bf M}}
\def\bm{{\bf m}}

\def\bt{{\bf t}}

\def\bV{{\bf V}}

\def\bX{{\bf X}}


\def\ib{\bar{i}}
\def\jb{\bar{j}}

\def\zb{\bar{z}}


\def\IB{\relax\hbox{$\inbar\kern-.3em{\rm B}$}}
\def\IC{{\bf C}}
\def\ID{\relax\hbox{$\inbar\kern-.3em{\rm D}$}}
\def\IE{\relax\hbox{$\inbar\kern-.3em{\rm E}$}}
\def\IF{\relax\hbox{$\inbar\kern-.3em{\rm F}$}}
\def\IG{\relax\hbox{$\inbar\kern-.3em{\rm G}$}}
\def\IGa{\relax\hbox{${\rm I}\kern-.18em\Gamma$}}
\def\IH{\relax{\rm I\kern-.18em H}}
\def\IK{\relax{\rm I\kern-.18em K}}
\def\IL{\relax{\rm I\kern-.18em L}}
\def\IP{\relax{\rm I\kern-.18em P}}
\def\IQ{\relax{{\vrule height1.5ex width.8pt depth0pt}\kern-.28em  Q\kern -.24em{\vrule height0.7em width.8pt depth0pt}}}
\def\IR{\relax{\rm I\kern-.18em R}}
\def\IZ{\relax\ifmmode\mathchoice{
\hbox{\cmss Z\kern-.4em Z}}{\hbox{\cmss Z\kern-.4em Z}}
{\lower.9pt\hbox{\cmsss Z\kern-.4em Z}} {\lower1.2pt\hbox{\cmsss
Z\kern-.4em Z}} \else{\cmss Z\kern-.4em Z}\fi}
\def\II{\relax{\rm I\kern-.18em I}}

\def\ndt{{\noindent}}


\def\CA{{\cal A}}

\def\CD{{\cal D}}
\def\CE{{\cal E}}

\def\CH{{\cal H}}

\def\CL{{\cal L}}
\def\CM{{\cal M}}
\def\CN{{\cal N}}
\def\CO{{\cal O}}

\def\CQ{{\cal Q}}
\def\CR{{\cal R}}

\def\CU{{\cal U}}
\def\CV{{\cal V}}
\def\CW{{\cal W}}

\def\CZ{{\cal Z}}

\def\ib{\bar{i}}
\def\jb{\bar{j}}

\def\zb{\bar{z}}


\def\Det{{\rm Det}}

\def\lime{{\rm Lim}_{\kern -16pt \vbox{\kern6pt\hbox{$\scriptstyle{\e \to 0}$}}}}

\def\naiveq{\qquad =^{\kern-12pt \vbox{\hbox{$\scriptscriptstyle{\rm naive}$}\kern5pt}} \qquad}

%
%

\hyphenation{anom-aly anom-alies coun-ter-term coun-ter-terms}
\def\tr{\, {\rm tr}\, }

\def\lspace{\ifx\answ\bigans{}\else\qquad\fi}
\def\lbspace{\ifx\answ\bigans{}\else\hskip-.2in\fi} 
\def\boxeqn#1{\vcenter{\vbox{\hrule\hbox{\vrule\kern3pt\vbox{\kern3pt
      \hbox{${\displaystyle #1}$}\kern3pt}\kern3pt\vrule}\hrule}}}
\def\mbox#1#2{\vcenter{\hrule \hbox{\vrule height#2in
          \kern#1in \vrule} \hrule}}  
%

\def\log{{\rm log}}

\def\darr#1{\raise1.5ex\hbox{$\leftrightarrow$}\mkern-16.5mu #1}


\def\half{{\textstyle{1\over2}}}

\def\roughly#1{\raise.3ex\hbox{$#1$\kern-.75em\lower1ex\hbox{$\sim$}}}


\def\inbar{\,\vrule height1.5ex width.4pt depth0pt}

\def\boxit#1{\vbox{\hrule\hbox{\vrule\kern8pt
\vbox{\hbox{\kern8pt}\hbox{\vbox{#1}}\hbox{\kern8pt}}
\kern8pt\vrule}\hrule}}
\def\mathboxit#1{\vbox{\hrule\hbox{\vrule\kern8pt\vbox{\kern8pt
\hbox{$\displaystyle #1$}\kern8pt}\kern8pt\vrule}\hrule}}


\def\mathcal#1{{\cal #1}}

\def\frac#1#2{{{ #1 }\over{ #2 }}}
\def\frac1#1{{1\over{#1}}}

\def\np#1#2#3{Nucl. Phys.\ {\bf B #1}(#2)#3}
\def\cmp#1#2#3{Comm. Math. Phys.\ {\bf #1}(#2)#3}
\def\plb#1#2#3{Phys. Lett. \ {\bf B #1} (#2) #3}  
\def\jhep#1#2#3{JHEP \ {\bf #1}(#2)#3}

\lref\Faltings{G.~Faltings, A proof for the Verlinde formula, J. Algebraic Geom. 3 (1994), no. 2, 347Ð374. MR 1257326 (95j:14013)}
\lref\Teleman{C.~Teleman, C.~Woodward,
``The Index Formula on the Moduli of G-bundles'', arXiv:math/0312154v4 [math.AG]}
\lref\LosevCR{
  A.~Losev, G.~W.~Moore, N.~Nekrasov and S.~Shatashvili,
Nucl.\ Phys.\ Proc.\ Suppl.\  {\bf 46}, 130 (1996).
[hep-th/9509151].
}
\lref\nsvz{
  V.~A.~Novikov, M.~A.~Shifman, A.~I.~Vainshtein and V.~I.~Zakharov,
Phys.\ Lett.\ B {\bf 217}, 103 (1989)..
}
\lref\niklect{N.~Nekrasov, ``Bethe states as defects in gauge theories'',
\vskip -0.15cm
\smallskip
\noindent
 {\tt http://scgp.stonybrook.edu/archives/8897}}

\lref\hitchin{N.~Hitchin, ``Stable bundles and integrable systems'', Duke Math
{\bf 54}  (1987) 91-114}
\lref\hid{N.~Hitchin, ``The self-duality equations on a Riemann surface'',
Proc. London Math. Soc. {\bf 55} (1987) 59-126 }
\lref\atbotti{M.~Atiyah, R.~Bott, 
``The Yang-Mills Equations Over
Riemann Surfaces'', Philosophical Transactions of the Royal Society of London. Series {\bA}, Mathematical and Physical Sciences, Volume 308, Issue 1505, pp. 523-615}

\lref\BaulieuNJ{
  L.~Baulieu, A.~Losev and N.~Nekrasov,
  ``Chern-Simons and twisted supersymmetry in various dimensions,''
Nucl.\ Phys.\ B {\bf 522}, 82 (1998).
[hep-th/9707174].
}
\lref\nikdiss{N.~Nekrasov, ``Four dimensional holomorphic theories'', PhD. thesis, Princeton University, 1996, UMI}

\lref\NekrasovUI{
  N.~A.~Nekrasov and S.~L.~Shatashvili,
  ``Quantum integrability and supersymmetric vacua,''
Prog.\ Theor.\ Phys.\ Suppl.\  {\bf 177}, 105 (2009).
[arXiv:0901.4748 [hep-th]].
}

\lref\NekrasovUH{
  N.~A.~Nekrasov and S.~L.~Shatashvili,
  ``Supersymmetric vacua and Bethe ansatz,''
Nucl.\ Phys.\ Proc.\ Suppl.\  {\bf 192-193}, 91 (2009).
[arXiv:0901.4744 [hep-th]].
}

\lref\NekrasovRC{
  N.~A.~Nekrasov and S.~L.~Shatashvili,
  ``Quantization of Integrable Systems and Four Dimensional Gauge Theories,''
[arXiv:0908.4052 [hep-th]].
}

\lref\gerasimovshatashvili{A.~Gerasimov, S.L.~Shatashvili, ``Higgs Bundles, Gauge Theories and Quantum Groups'', \cmp{277}{2008}{323-367}, arXiv:hep-th/0609024}

\lref\gstwo{A.~Gerasimov, S.L.~Shatashvili, ``Two-dimensional gauge theories and quantum integrable systems'', arXiv:0711.1472, in, {\it "From Hodge Theory to Integrability and TQFT: tt*-geometry"}, pp. 239-262,
R.~Donagi and K.~Wendland, Eds., Proc. of Symposia in
Pure Mathematics Vol. 78, AMS, 
Providence, Rhode Island}
\lref\NekrasovXDA{
  N.~Nekrasov, V.~Pestun and S.~Shatashvili,
[arXiv:1312.6689 [hep-th]].
}

\lref\gntd{A.~Gorsky, N.~Nekrasov, ``Hamiltonian systems of Calogero type and two dimensional Yang-Mills theory'', arXiv:hep-th/9304047, \np{414}{1994}{213-238}}

\lref\mnph{J.A.~Minahan, A.P.~Polychronakos, ``Interacting Fermion Systems from Two Dimensional QCD,'' 
\plb{326}{1994}{288-294}, arXiv:hep-th/9309044    \semi
~~~~``Equivalence of Two Dimensional QCD and the $c=1$ Matrix Model'', \plb{312}{1993}{155-165}, arXiv:hep-th/9303153   \semi
~~~~ ``Integrable Systems for Particles with Internal Degrees of Freedom,''
\plb{302}{1993}{265-270}, arXiv:hep-th/9206046}

    \lref\korepin{V.~Korepin, N.~Bogolyubov, A.~Izergin, ``Quantum Inverse Scattering Method and Correlation Functions'', Cambridge  Monographs on Mathematical Physics, Cambridge University Press, 1997}
\lref\dubrovintt{B.~Dubrovin, ``Geometry and Integrability of Topological-Antitopological Fusion'', arXiv:hep-th/9206037,   \cmp{152}{1993}{539-564} }

\lref\WittenZZ{
  E.~Witten,
  ``Mirror manifolds and topological field theory,''
In *Yau, S.T. (ed.): Mirror symmetry I* 121-160.
[hep-th/9112056].
}

\lref\mz{J.~Minahan, K.~Zarembo, ``The Bethe-Ansatz for ${\CN}=4$ Super Yang-Mills, ''
arXiv:hep-th/0212208 , \jhep{0303}{2003}{013}}
\lref\vafacr{C.~Vafa, ``Topological Mirrors and Quantum Rings,'' in, {\it Essays on Mirror Manifolds}, ed. S.-T.~Yau (Intl.Press, 1992)}
\lref\vc{S.~Cecotti, C.~Vafa, ``Topological Anti-Topological Fusion'', \np{367}{1991}{359-461}}
\lref\gnell{A.~Gorsky, N.~Nekrasov, ``Elliptic Calogero-Moser system from two dimensional current algebra'', arXiv: hep-th/9401021}
\lref\gnthd{A.~Gorsky, N.~Nekrasov, ``Relativistic Calogero-Moser model as gauged WZW theory'', \np{436}{1995}{582-608}, arXiv:hep-th/9401017}
  
\lref\nazarov{M.~Nazarov, V.~Tarasov,``On irreducibility of tensor products of Yangian modules associated with skew Young diagrams'', arXiv:math/0012039, Duke~Math.~J. {\bf 112} (2002), 343-378\semi
M.~Nazarov, V.~Tarasov, ``On Irreducibility of Tensor Products of Yangian Modules'',
arXiv:q-alg/9712004,
Internat.~Math.~Research~Notices (1998) 125-150\semi
M.~Nazarov, V.~Tarasov, ``Representations of Yangians with Gelfand-Zetlin Bases'',
arXiv:q-alg/9502008,  J.~Reine~Angew.~Math. {\bf 496} (1998) 181-212\semi
M.~Nazarov, V.~Tarasov, ``Yangians and Gelfand-Zetlin bases'', arXiv:hep-th/9302102, Publ.~Res.~Inst.~Math.~Sci. Kyoto {\bf 30} (1994) 459-478}  
  
\lref\WitDonagi{R.~ Donagi, E.~ Witten,
``Supersymmetric Yang-Mills Theory and
Integrable Systems'', hep-th/9510101, Nucl.Phys.{\bf B}460 (1996) 299-334}
\lref\Witfeb{E.~ Witten, ``Supersymmetric Yang-Mills Theory On A
Four-Manifold,'' J. Math. Phys. {\bf 35} (1994) 5101.}
 
\lref\WittenTwoD{E.~Witten, ``Two dimensional gauge theory revisited'', arXiv:hep-th/9204083}

\lref\gmmm{A.~Gorsky, A.~Marshakov, A.~Mironov, A.~Morozov, ``${\CN}=2$ Supersymmetric QCD and Integrable Spin Chains: Rational Case $N_f < 2N_c$'', arXiv: hep-th/9603140, \plb{380}{1996}{75-80}}
\lref\ggm{A.~Gorsky, S.~Gukov, A.~Mironov,``SUSY field theories, integrable systems and their stringy/brane origin -- II'', arXiv:hep-th/9710239,
\np{518}{1998}{689-713}\semi
A.~Gorsky, S.~Gukov, A.~Mironov,
``Multiscale ${\CN}=2$ SUSY field theories, integrable systems and their stringy/brane origin -- I
'', arXiv:hep-th/9707120, \np{517}{1998}{409-461} \semi
R.~Boels, J.~de Boer,
``Classical Spin Chains and Exact Three-dimensional Superpotentials'',
arXiv:hep-th/0411110}

\lref\BlThlgt{M.~ Blau and G.~ Thompson, ``Lectures on 2d Gauge
Theories: Topological Aspects and Path
Integral Techniques", hep-th/9310144.}

\lref\kolyaf{N.~Reshetikhin, ``The functional equation method in the theory of exactly soluble quantum systems'', ZhETF {\bf 84} (1983), 1190-1201 (in Russian) Sov. Phys. JETP {\bf 57} (1983), 691-696 (English Transl.)}

\lref\baxteriii{R.J.~Baxter, ``Partition Function of the Eight-Vertex Lattice Model'' , Ann. Phys. {\bf 70} (1972) 193-228\semi
``One Dimensional Anisotropic Heisenberg Chain'', Ann. Phys. {\bf 70} (1972) 323-337\semi ``Eight-Vertex Model in Lattice Statistics and One Dimensional Anisotropic Heisenberg Chain I, II, III'', Ann. Phys. {\bf 76} (1973) 1-24, 25-47, 48-71}

\lref\baekr{
A.~N.~Kirillov,
N.~Yu.~Reshetikhin, ``Representations of Yangians
and multiplicities of the inclusion of the irreducible components
of the tensor product of representations of simple Lie algebras'' ,
(Russian) Zap. Nauchn. Sem. Leningrad. Otdel. Mat. Inst. Steklov.
(LOMI) 160 (1987), Anal. Teor. Chisel i Teor. Funktsii. 8,
211--221, 301; translation in J.~Soviet~Math. {\bf 52} (1990), no. 3,
3156--3164}

\lref\wittgr{E.~ Witten, ``The Verlinde Algebra And The Cohomology Of
The Grassmannian'',  hep-th/9312104}

\lref\niksw{N.~Nekrasov, ``Seiberg-Witten prepotential
from instanton calculus,'' arXiv:hep-th/0206161,  arXiv:hep-th/0306211}

\lref\nikokounkov{N.~Nekrasov, A.~Okounkov, ``Seiberg-Witten theory and random partitions,''
 arXiv:hep-th/0306238}

\lref\baxter{R.~Baxter, ``Exactly solved models in statistical mechanics'', London, Academic Press, 1982}

\lref\issues{A.~Losev, N.~Nekrasov, S.~Shatashvili, ``Issues in topological gauge theory'', \np{534}{1998}{549-611}, 
arXiv:hep-th/9711108}

\lref\gerasimov{A.~Gerasimov, ``Localization in
GWZW and Verlinde formula,'' hepth/9305090}

\lref\btverlinde{M.~ Blau, G.~Thomson,
``Derivation of the Verlinde Formula from Chern-Simons Theory and the
$G/G$ model,'' \np{408}{1993}{345-390} }

\lref\gyangian{H.~Nakajima, ``Quiver varieties and finite dimensional representations of quantum affine algebras,'' 
arXiv:math/9912158}
\lref\mvyangian{M.~Varagnolo,  ``Quiver varieties and Yangians,'' arXiv:math/0005277}

\lref\yangyang{C.~N.~Yang, C.~P.~Yang, J.~Math.~Phys~{\bf 10} (1969) 1115}

\lref\nikfive{N.~Nekrasov, ``Five dimensional gauge theories and relativistic integrable systems'',
\np{531}{1998}{323-344},
arXiv:hep-th/9609219}

\lref\phases{E.~Witten, ``Phases of ${\CN}=2$ Theories in Two Dimensions",
Nucl. Phys. {\bf B403} (1993) 159, hep-th/9301042}

\lref\nekhol{N.~Nekrasov, ``Holomorphic bundles and many-body systems'', arXiv:hep-th/9503157, \cmp{180}{1996}{587-604} }  
    
\Title{ \vbox{\baselineskip12pt
\hbox{}
\hbox{TCD-MATH-14-06}
\hbox{HMI-14-01}
\hbox{}}}
{\vbox{
\bigskip
\bigskip
 \centerline{BETHE/GAUGE CORRESPONDENCE}
 \bigskip
 \centerline{ON CURVED SPACES}
}}
\medskip
\centerline{\authorfont Nikita
Nekrasov\footnote{$^{\, a}$}{On leave of absence
from IHES, Bures-sur-Yvette, France and IITP and ITEP, Moscow, Russia}$^{,1}$,
and Samson  Shatashvili\footnote{$^{\, b}$}{On leave of absence
from Euler International Mathematical Institute, Saint-Petersburg, Russia and IITP, Moscow, Russia}$^{,1,2,3}$}
\vskip 0.5cm
\centerline{\it $^{1}$ Simons Center for Geometry and Physics, Stony Brook NY 11794-3636 USA}
\centerline{\it $^{2}$ Hamilton Mathematical Institute, Trinity College,
Dublin 2, Ireland}
\centerline{\it $^{3}$ School of Mathematics, Trinity College, Dublin 2, Ireland}
\vskip 0.1cm

\bigskip
\ndt
{\abstractfont
Bethe/gauge correspondence identifies supersymmetric vacua of massive gauge theories invariant under the two dimensional ${\CN}=2$ Poincare supersymmetry with the stationary states of some quantum integrable system.  The supersymmetric theory can be twisted in a number of ways, producing a topological field theory.  For these theories we compute the handle gluing operator $\CH$. We also discuss  the Gaudin conjecture on the norm of Bethe states and its connection to $\CH$.}
\vfill\eject
\newsec{Introduction}

The correspondence between supersymmetric gauge theory and quantum integrable system was discovered, in an example, in \ref\higgs{G.~Moore, N.~Nekrasov, S.~Shatashvili,
``Integration over the Higgs branches'', 
\cmp{209}{2000}{97-121}, arXiv:hep-th/9712241}, further explored in \gerasimovshatashvili,\gstwo,  where it was speculated that such a correspondence should be a general property of a larger class of gauge theories in various space-time 
 dimensions, and finally formulated and explained in \NekrasovUI\NekrasovUH\NekrasovRC\ . Prior to \higgs\  a connection of two dimensional pure Yang-Mills theory with massive matter to spin and many-body systems systems with long-range interaction 
 appeared in
  \gntd,\mnph\ (the embedding of pure two dimensional Yang-Mills theory into the supersymmetric theory was discussed in \WittenTwoD). The three dimensional lift of that gauge theory describes the relativistic interacting particles \gnthd, while the four dimensional theories lead to elliptic generalizations
\gnell. 

In \NekrasovUI\NekrasovUH\NekrasovRC\  the correspondence, called the {\it Bethe/gauge correspondence}, between the two dimensional ${\CN}=2$ supersymmetric gauge theories and quantum integrable systems was formulated 
in full generality. The novelty of this correspondence is the identification of the Planck constant of the quantum integrable system with a twisted mass parameter on the gauge theory side. 
The dictionary further identifies the Yang-Yang function of the quantum integrable system with the effective twisted superpotential ${\tilde W}^{\rm eff}$ of the gauge theory, which is a function  of the Coulomb moduli, taking values in the Cartan 
subalgebra of the Lie algebra of the gauge group of the theory. 
The twisted superpotential can be computed exactly (given by one-loop formula).  It allows to describe all the supersymmetric vacua of physical gauge theory in finite volume, i.e. on the space-time with the cylinder geometry, assuming all the matter fields are massive. The vacua correspond to the generalized critical points of
${\tilde W}^{\rm eff}$, which are the critical 
points of the shifted effective twisted superpotential 
\eqn\wsh{{\tilde W}^{\rm eff}({\sigma}_{1},\ldots , {\sigma}_{N})-2 \pi {\ii} \sum_{a=1}^{N}  n_{a}\sigma_{a}} 
In Bethe/gauge correspondence these equations are Bethe equations of corresponding quantum integrable system. In \NekrasovUI\NekrasovUH\NekrasovRC\ explicit formulas for ${\tilde W}^{\rm eff}$ for a very large class of theories were given, including the properly two dimensional theories, three dimensional theories compactified on a circle, and four dimensional theories compactified on the two-torus or subject to the two dimensional $\Omega$-deformation \niksw; the interested reader can consult these papers for details. Our notations follow mostly the conventions from these papers. 

The effective action of the gauge theory contains, in addition to the twisted F-terms determined by ${\tilde W}^{\rm eff}$, the effective dilaton interaction ${\tilde U}^{\rm eff}({\sigma}) {\CR}^{(2)}$, describing the response of the theory to the background gravitational field.

The effective gravitational interaction and its r\^ole in the Bethe/gauge-correspondence is the subject of the present short note.

\newsec{The gauge theory}

Before we turn to the topologically twisted theories, we give a brief review of the relevant physical gauge theories,  following the notations of \NekrasovUI, \NekrasovUH.

\subsec{Gauge theories with four supercharges}

We study two dimensional ${\CN} = (2,2)$ supersymmetric gauge theory
 with some matter. The matter fields are generally in
the chiral multiplets which we denote by the letters ${\bf Q}$, ${\tilde{\bf Q}}$, and ${\bf \Phi}$  (sometimes we use $\bX$ to denote matter fields without reference to their gauge representation type), the gauge fields are in the vector multiplet $\bf V$. We also use the twisted chiral multiplets $\bf\Sigma$, as e.g. the field strength ${\bf\Sigma} = {\CD}_{+}{\bar \CD}_{-}{\bV}$ is in the twisted chiral multiplet.
\eqn\multiplet{\eqalign{& {\bV}={\t}^{-}
{\bar\t}^{-}(A_{0}-A_{1})+{\t}^{+}{\bar\t}^{+}(A_{0}+A_{1})-\sqrt{2}{\s}{\t}^{-}{\bar\t}^{+}
-\sqrt{2}{\bar\s}{\t}^{+}{\bar\t}^{-}+ \cr
&
+2i{\t}^{-}{\t}^{+}({\bar\t}^{-}{\bar\l}_{-} +
{\bar\t}^{+}{\bar\l}_{+}) + 2 i {\bar\t}^{+}{\bar \t}^{-}({\t}^{+}{\l}_{+} +{\t}^{-}{\l}_{-})+2{\t}^{-}{\t}^{+}{\bar \t}^{-}{\bar\t}^{+}H \ , \cr}}
where we use a notation $H$ for the auxiliary field (in most textbooks it is denoted by $D$).
\eqn\chirl{{\bX} = X ( y ) + \sqrt{2} \left( {\t}^{+} {\psi}_{+}(y)  + {\t}^{-} {\psi}_{-}(y)  \right) + {\t}^{+}{\t}^{-} F(y)}
where $$
y^{\pm} = x^{\pm} - i{\t}^{\pm}{\bar\t}^{\pm}\ ,
$$
and the twisted chiral multiplet $\bf\Sigma$:
\eqn\sig{{\bf\Sigma} =  {\s}({\tilde y}) +i\sqrt{2}\left( {\t}^{+}{\bar \l}_{+}({\tilde y}) -{\bar \t}^{-}{\l}_{-}({\tilde y}) \right)  + \sqrt{2}{\t}^{+}{\bar \t}^{-} \left(  H ({\tilde y}) -i F_{01} \right)}
where $F_{01} = {\p}_{0}A_{1} - {\p}_{1}A_{0} + [A_{0}, A_{1}]$ is the gauge field strength, and
\eqn\ypmv{
{\tilde y}^{\pm} = x^{\pm} \mp i {\t}^{\pm}{\bar\t}^{\pm}}

\subsubsec{Lagrangians}

The action of the
corresponding two dimensional quantum field theory is a sum of three types of terms - the $D$-terms, the $F$-terms and the twisted $F$-terms:
\eqn\dff{\eqalign{
A=\int {\rm d}^{2}x\, {\rm d}^{4}{\theta} ({\tr} \left( {\bf\Sigma}{\bf\bar \Sigma} \right) + &{\bK} ( e^{{\bV}/2}\, {\bX}\ , \,{\bar {\bX}}\, e^{{\bV}/2}))
+\int {\rm d}^{2}x\, {\rm d}{\theta}^{+}{\rm d}{\t}^{-} ( W({\bX}) \, + \, {\rm c.c.})\cr
&+\int {\rm d}^{2}x\, {\rm d}{\t}^{+}{\rm d}{\bar\t}^{-} ({\tilde W}({\bf\Sigma}) \, + \, {\rm c.c.})}}

\subsubsec{Global symmetries and twisted masses}

The typical ${\CN}=(2,2)$ gauge theory has the matter fields $\bX$ transforming in some
linear representation $\CR$ of the gauge group $G$.
Let us specify the decomposition of $\CR$ onto the irreducible representations of $G$:
\eqn\CRdc{{\CR} = \bigoplus_{\bi} \, {\bM}_{\bi} \otimes R_{\bi}}
where $R_{\bi}$ are the irreps of $G$, and ${\bM}_{\bi}$ are the multiplicity spaces.
The group
\eqn\glbgrp{G_{\rm f}^{\rm max} = \times_{\bi} \, U({\bM}_{\bi})}
acts on $\CR$ and this action commutes with the gauge group action. The actual global symmetry
group $G_{\rm f}$ of the theory may be smaller then \glbgrp : $G_{\rm f} \subset G_{\rm f}^{\rm max}$, as it has to preserve both $D$ and the $F$-terms in the action.

The theory we are interested in can be deformed by turning on the so-called twisted masses
${\tilde m}$ \ref\agf{L.~Alvarez-Gaume, D.Freedman, \cmp{91}{1983}{87}}, which belong to the complexification of the Lie algebra of the maximal torus of $G_{\rm f}$:
\eqn\twmsss{{\tilde m} = \left( {\tilde m}_{\bi}  \right) \, , \, {\tilde m}_{\bi} \in {\rm End} \left( {\bM}_{\bi} \right) \cap G_{\rm f}}
The superspace expression for the twisted mass term is \ref\sstwm{S.J.~Gates, \np{238}{1984}{349}},
\ref\sstwmii{S.J.~Gates, C.M.~Hull,  M.~Rocek, \np{248}{1984}{157}}:
\eqn\twmass{{\CL}_{\widetilde{\rm mass}} = \int {\rm d}^{4}{\theta}
\ {\tr}_{\CR} \, {\bX}^{\dagger} \left( \sum_{\bi} e^{{\tilde\CV}_{\bi}}\otimes {\rm Id}_{R_{\bi}}\right) {\bX} }
where
\eqn\tvm{{\tilde\CV}_{\bi} = {\tilde m}_{\bi}\, {\theta}_{+}{\bar\theta}_{-}}
The twisted masses which preserve the ${\CN}=4$ supersymmetry will be denoted by $\m$, and
the ones which break it down to
${\CN}=2$, by $u$.

\subsec{Microscopic theories}

Assuming the theory is two dimensional at high energies, ${\Lambda} \to \infty$, the two-observable part of the topological theory action is determined by the theta angle and the Fayet-Illiopoulos term of the microscopic theory:
\eqn\cwl{{\tilde W}^{\rm uv} = \sum_{i} t_{i} \, {\tr}_{N_{i}} {\sigma}_{i}}
where we assumed the gauge group to be the product of the $U(N_i)$ factors, and
\eqn\tcm{t_{i} = {{\vartheta}_{i}\over 2\pi} + {\ii} r_{i}}

Finally, if the microscopic theory is a four dimensional ${\CN}=2$ theory, with complexified couplings ${\tau}_{i}$, 
\eqn\tui{{\tau}_{i} = {{\vartheta}_{i} \over 2\pi} + {4\pi \ii \over g_{i}^{2}}} 
subject to a two-dimensional $\Omega$-deformation with the parameter $\ve$, then the two-observable is given by:
\eqn\cwfd{{\tilde W}^{\rm uv}  = \sum_{i} {2{\pi} {\tau}_{i}\over \ve}  \, {\tr}_{N_{i}} {\sigma}_{i}^2}

\subsec{Effective theory}

When the twisted masses are turned on in  the generic fashion, the matter fields
are massive and can be integrated out. As a result, the theory becomes an effective pure ${\CN}=2$ gauge theory with an infinite
number of interaction terms in the Lagrangian, with the high derivative terms suppressed by the inverse masses of the fields we integrated out. Of all these terms the $F$-terms, i.e. the effective superpotential, or the twisted $F$-terms, i.e. the effective twisted superpotential, can be computed exactly. In fact, these terms only receive one-loop contributions. Let $\tilde m$ denote collectively the set of the twisted masses of the fields we are
integrating out.
One can write:
\eqn\effsp{\widetilde{W^{\rm eff}} = {\tilde W}^{\rm uv} + 
{\tilde W}_{\rm matter} + {\tilde W}_{\rm gauge} }
The subscript ``matter'' in ${\tilde W}_{\rm matter}$ in \effsp\  stresses the fact that it only includes the loop contribution of the matter fields:
\eqn\wmat{{\tilde W}_{\rm matter} = 
{d\over ds} \Biggr\vert_{s=0} {{\Lambda}_{\rm uv}^{s} \over {\Gamma}(s)} \int_{0}^{\infty} {dt\over t^{2}}t^{s}\  {\tr}_{\CR} \, e^{-t \left( {\sigma} + {\tilde m} \right)}}
If the theory is properly two dimensional, that is $\CR$ is finite dimensional, then \wmat\ can be evaluated to give:
\eqn\wmatii{{\tilde W}_{\rm matter}^{2d} = {\tr}_{\CR} \, \left( {\sigma} + {\tilde m} \right) \left( {\log} 
\left( {{\sigma} + {\tilde m} \over {\Lambda}_{\rm uv}} \right) - 1 \right)}
The ultraviolet scale $\Lambda_{\rm uv}$ is absorbed in the renormalized Fayet-Illiopoulos term. 

{} If the theory is three dimensional, compactified on a circle of radius $R$, then the representation $\CR$ is the space of periodic functions $\psi(z)$,  $z \sim z + 2\pi R$, valued in some finite dimensional representation ${\CR}_{3}$ of the gauge group $G$. The trace ${\tr}_{\CR}$ involves the sum over the Kaluza-Klein modes, ${\tilde m} \to {\tilde m} + {{\ii} n \over R}$, $n \in {\IZ}$. 
The sum forces the ``proper-time'' $t$ in \wmat\ to be quantized:
  $t = 2\pi k R$, $k \in {\IZ}_{+}$, leading to:
  \eqn\wmatiii{{\tilde W}_{\rm matter}^{3d} = {\tr}_{{\CR}_{3}} \, {\rm Li}_{2} \left( e^{- 2\pi R 
( {\sigma} + {\tilde m} )} \right)}
In \NekrasovUI\NekrasovUH\NekrasovRC\ one can find the expressions for the effective twisted superpotentials of the  four dimensional ${\CN}=1$ theory, compactified on a torus $T^2$ or 
the four dimensional ${\CN}=2$ theory subject to a special $\Omega$-background. These formulas are much more complicated and we shall not write them here.

There are other massive fields which can be integrated out on the Coulomb branch. For example, the ${\bg}/{\bt}$-components of the vector multiplets (where $\bg$ denotes Lie  algebra  corresponding to Lie groups $G$ and $\bt$ is its Cartan sub-algebra), the $W$-bosons
and their superpartners. Their contribution to the effective twisted superpotential is rather simple: in two dimensions
\eqn\gsup{
{\tilde W}_{\rm gauge}^{2d} = 
\sum_{{\a} \in {\Delta}} \, \langle
{\a} , {\s} \rangle \, \left( \,
{\rm log} \, \langle
{\a} , {\s} \rangle \, - 1\, \right)
= - 2{\pi} i \, \langle {\rho} , {\s} \rangle}
where
\eqn\hlfsm{{\rho} = {\half} \sum_{{\a} \in {\Delta}_{+}} {\a}}
is half the sum  of the positive roots of $\bg$.
In three dimensions,
\eqn\gsupiii{
{\tilde W}_{\rm gauge}^{3d} = 
\sum_{{\a} \in {\Delta}} \, {\rm Li}_{2} \left( e^{\langle
{\a} , {\s} \rangle} \right) 
= - 2{\pi} i \, \langle {\rho} , {\s} \rangle - {h_{\bg}^{\vee} \over 24} \langle {\sigma}, {\sigma} \rangle}
It may appear that the expressions \gsup\gsupiii\ are inconsistent
with the gauge invariance, however the effective interaction \gsup\ is gauge invariant (in the three dimensional case this has to do with the effective Chern-Simons term and the celebrated shift $k \to k + h_{\bg}^{\vee}$, see also the Eq. (3.7) below).

\subsubsec{Superpotential deformations and twisted masses}

The supersymmetric field theories also have the superpotential deformations, which correspond to the $F$-terms in \dff. The  superpotential $W$ has to be a holomorphic gauge invariant
function of the chiral fields, such as ${\Phi}, Q, {\tilde Q}$.
It may be not invariant under the maximal symmetry group $G_{\rm f}^{\rm max}$,
thus breaking it to a subgroup $G_{\rm f}$. 
For example, the so-called {\it complex mass} of the fundamental and anti-fundamental fields comes from
the superpotential $
W_{m}  = \sum_{a,b}m_{a}^{b} {\tilde Q}_{b}Q^{a}
$,
which breaks the $U(n_{\bf f}) \times U(n_{\overline{\bf f}})$
group down to $U(1)^{{\rm min}(n_{\bf f}, n_{\overline{\bf f}})}$.

{}In all cases discussed in this paper, in spacetime dimensions two, three and four, one can consider more sophisticated superpotentials,  involving  the fundamental, anti-fundamental, and adjoint chiral fields:
\eqn\wqfq{
W_{m,s}  = \sum_{a,b} {\tilde Q}^{a}m_{a}^{b}(\Phi)Q_{b} = \sum_{a,b; s}\, m_{a; s}^{b} {\tilde Q}^{a}
{\Phi}^{2s}Q_{b}}

\newsec{Effective action and topological theory}

The supersymmetric theory can be topologically twisted, in a number of ways, so as to preserve some supersymmetry even on the curved backgrounds.

This is accomplished by identifying the $U(1)$ Lorentz symmetry with a diagonal subgroup in the product of the Lorentz $U(1)$ and one of the R-symmetry $U(1)$'s.  The vector multiplet fields become, after the twist, 
a gauge field $A$, a one-form fermion $\lambda$ in the adjoint representation, two scalar
fermions $\eta, \chi$ both in the adjoint, and the complex scalar ${\sigma}, {\bar\sigma}$. 

The twist in the matter sector is defined ambiguously, 
the choice being the additional twist in the flavor group. The only requirement is that the superpotential $W$ after the twist becomes a $(1,0)$-form on the world sheet $\Sigma$. This is always possible to achieve if the superpotential is quasi-homogeneous with respect to some R-symmetry.

Let us denote the topological supercharge, i.e. the supercharge of the original theory, which becomes, upon twisting, a Lorentz scalar, by $\CQ$:
\eqn\qcha{{\CQ}^{2} = 0} 
One can associate three topological theories to the given physical supersymmetric gauge theory with a choice of twist. Namely, the original microscopic theory, the effective gauge theory, which is obtained from the microscopic theory by integrating out the matter multiplets, and the effective abelian gauge theory, which is obtained from the effective gauge theory by integrating our the massive vector multiplets. The advantage of the topological field theory formalism is the possibility of throwing away the non-local terms, generated by the higher loops of the integrated out fields, as long as one can show that these are $\CQ$-exact.

At every scale $\Lambda$ the action functional of the resulting topological theory can be written as:
\eqn\twod{S^{\Lambda} = \int_{\Sigma} \left( {\partial {\CW}^{\Lambda} \over {\partial {\sigma}^{a}}} F^{a} + {\half} {\partial^2 {\CW}^{\Lambda} \over {\partial {\sigma}^{a}\partial {\sigma}^{b}}} {\psi}^{a} \wedge {\psi}^{b} + {\CU}^{\Lambda}({\sigma}) {\CR}^{(2)} \right) + {\CQ} \left( {\ldots} \right) }
which is the sum of the two-observable constructed out of some gauge-invariant (modulo little subtleties) function ${\CW}^{\Lambda}({\sigma})$ of  the complex adjoint scalar $\sigma$ in the vector multiplet, and the zero-observable ${\CU}^{\Lambda}({\sigma})$, integrated against the Euler density
${\CR}^{(2)}$. The scalar fields $(\sigma^{a})$ parametrize the 
Lie algebra of the gauge group for $\Lambda$ above the scale of the mass of the $W$-bosons. For $\Lambda$ below that scale the scalars $({\sigma}^{a})$ parametrize the Cartan sublagebra of the gauge group. The two-observable part of the topological theory action is directly related to the effective action of the physical theory:
\eqn\www{{\CW}^{\Lambda}({\sigma}) = {\tilde W}^{\rm eff} ({\sigma}; {\Lambda})}
where in the right hand side we made explicit the scale dependence. 

We have discussed the ultraviolet expressions for ${\tilde W}^{\rm uv} = {\CW}^{\infty}$,   in the previous section. There is one subtlety in the case the theory is the three dimensional theory compactified to two dimensions on a circle. 

Indeed, suppose the microscopic theory is three dimensional, with ${\CN}=2$ supersymmetry. It cannot be topologically twisted on a general three-manifold. However, on manifolds with an isometry, it has a twist, producing a supercharge $\CQ$ which squares to the translation along the isometry.  The theory can be deformed by Chern-Simons couplings
\eqn\csk{S_{\rm CS} = \sum_{i} {k_{i} \over 4\pi} \int {\tr}_{N_{i}} \left( A_{i} \wedge dA_{i} + {2\over 3} A_{i} \wedge A_{i} \wedge A_{i} \right)}
which can be completed to a $\CQ$-closed interaction if accompanied by the term \BaulieuNJ:
\eqn\fif{\sum_{i} k_{i} \int {\tr}_{N_{i}} \left( {\varphi}_{i}F_{A_{i}} + {\lambda}_{i}{\lambda}_{i} \right) \wedge dz}
Now suppose the theory is compactified on a circle $S^{1}$ of radius $R$, $z \sim z + 2\pi R$, 
then the two dimensional field $\sigma_{i}$ is  the sum of the real adjoint scalar ${\varphi}_{i}$ and the component $\iota_{{\partial}_{z}}  A_{i}$ of the gauge field along the circle
\eqn\sifa{{\sigma}_{i} = \iota_{{\partial}_{z}} A_{i} + {\ii} {\varphi}_{i}}
In this case the two-observable part of the topological theory action is determined by the Chern-Simons levels: 
\eqn\cwcs{{\tilde W}^{\rm uv}  = \sum_{i} 2{\pi} k_{i} R \, {\tr}_{N_{i}} {\sigma}_{i}^2}

\subsec{BPS equations}

The path integral of the topological field theory obtained by the twist of the gauge theory reduces to the finite-dimensional
integral over the moduli space ${\CM}_{\Sigma}$ of solutions to a system of partial differential equations over the Riemann surface $\Sigma$:
\eqn\bpseq{\eqalign{& D_{\zb} X^{i} = - G^{i\ib} {{\partial}W_{\zb}^{*} \over {\partial}{\bar X}^{\ib}} \cr
& F_{z\zb} + {\mu} (X, {\bar X}) = 0 \cr}}
considered up to the action of gauge group.
 
Here $X^i$ denote the holomorphic coordinates on the space of scalar components of the chiral matter multiplets, and
\eqn\kmet{
G_{i\jb} = {{\partial}^{2}{\bf K} (X, {\bar X}) \over {\partial}{X^{i}}{\partial}{{\bar X}^{\jb}}} }
is the K\"ahler metric defined by the kinetic term \dff, ${\mu}(X, {\bar X})$ is the ${\bg}^{*}$-valued moment map  including the Fayet-Illiopoulos terms, associated with the symplectic form
\eqn\sympl{{\omega} = \sum_{i, {\jb}} G_{i\jb} dX^{i} \wedge d{\bar X}^{\jb}}
Finally,  $W_{z}$
denotes the superpotential $W({\bf X})$ (not to be confused with the twisted superpotential), cf. \dff.  The subscript $z$ in $W_z$ signifies the fact that upon the $A$-type twist, the kind of twist we use in this paper, the superpotential becomes a $(1,0)$-form on the world sheet. 

For example, in the theory with the ${\CN}=4$ fundamental hypermultiplet $(Q, {\tilde Q})$ and the adjoint chiral multiplet $\Phi$, with the superpotential \wqfq\ 
\eqn\qfqs{W = {\tilde Q}{\Phi}^{2s} Q}
the symmetric twist makes $\Phi$ a $(1,0)$ form, while $(Q, {\tilde Q})$ become the $({\half} - s, 0)$-forms. The more general twists
make ${\Phi}$ a $(u,0)$-differential, the $Q$ an $({\half} - s{\cdot} u+t,0)$-differential, and ${\tilde Q}$ an $({\half} - s{\cdot} u -t,0)$-differential. As a result, the superpotential \qfqs\ is the $(1,0)$-differential. 

The moduli space \bpseq\ can be deformed, by scaling the superpotential down to zero, to the moduli space of stable Higgs pairs $({\CE}, {\bX})$, where $\CE$ is a holomorphic $G_{\IC}$-bundle on $\Sigma$, and $\bX$ is the holomorphic section of the vector bundle 
\eqn\cre{\CR_{\CE, \rho} = \oplus_{\bi} {\bM}_{{\rho}, \bi} \otimes {\CR}_{\bi}   } associated with $\CE$ and the representations ${\CR}_{\bi}$ of $G$, and the homomorphism $\rho$ from $U(1)_{\Sigma}$ to $G_{\rm f}$.
For the example \qfqs\ the bundle $\CR_{\CE, \rho}$ is the sum
\eqn\qrqfs{
\CR_{\CE, \rho} = \left( E \otimes E^{*} \otimes K_{\Sigma}^{u} \right)\oplus \left( E \otimes K_{\Sigma}^{ - su - t + {\half}} \right) \oplus \left( E^{*} \otimes K_{\Sigma}^{ - su + t + {\half}} \right)}
where $E$ is a holomorphic rank $N$ vector bundle on $\Sigma$.
 
The moduli space $\CM$ is acted upon by the subgroup $G_{\rm f}^{\rho}$ of $G_{\rm f}$, commuting with the image of $\rho$. Generically, $G_{\rm f}^{\rho}$ has the same maximal tori as 
$G_{\rm f}$. Let us denote it by $T_{\rm f}$. Let us denote the image of the canonical generator of $U(1)_{\Sigma}$ in the corresponding Cartan subalgebra $LieT_{\rm f}$ by $S$. 

The path integral of the topological field theory computing the correlation functions of the $\CQ$-invariant observables reduces to the $T_{\rm f}$-equivariant integral of the corresponding
cohomology classes on $\CM$:
\eqn\corrls{\langle {\CO}_{1} \ldots {\CO}_{p} \rangle = 
\int_{\CM} {\Omega}_{1} \wedge \ldots \wedge {\Omega}_{p}}

\subsec{Effective action  from the index theorem}

The effective twisted superpotential ${\CW}$ and the effective dilaton $\CU$, determining the topological theory action \twod\
can be computed from the equivariant Atiyah-Singer, or Riemann-Roch-Grothendieck index theorem
\nikdiss. 

\subsubsec{The matter contribution}
The idea is to consider the two maps: the projection 
\eqn\pbun{p: {\CM} \to Bun_{G_{\IC}}}
which sends the pair $({\CE}, {\bX})$ to $\CE$, and the inclusion
\eqn\ibun{i: Bun_{G_{\IC}} \to {\CM}^{\circ}}
which sends the bundle ${\CE}$ to the pair $({\CE},0)$. The latter need not be stable, therefore the target ${\CM}^{\circ}$ of that map is somewhat larger then $\CM$, and is, likely, singular. We shall ignore this subtlety in what follows. 

The cohomology classes \corrls\ get pushed forward to $Bun_{G_{\IC}}$:
\eqn\pushf{p_{*}( {\Omega}_{1} \wedge \ldots \wedge {\Omega}_{p} ) = {i^{*} ({\Omega}_{1} \wedge \ldots \wedge {\Omega}_{p} ) \over
Euler ( {\CN} )}}
where  $\CN$ is the normal bundle to the locus $\bX = 0$ in $\CM$ and {\it Euler} denotes the equivariant Euler class. 

From \pushf\ we derive:
\eqn\asth{\eqalign{& \int_{\Sigma} {\CO}_{{\CW}_{\rm matter}}^{(2)} + c_{1}({\Sigma})  {\CO}_{{\CU}_{\rm matter}}^{(0)} = \cr
& \qquad
{d\over ds}\Biggr\vert_{s=0} {{\Lambda}_{\rm uv}^{s}\over {\Gamma}(s)} \int_{0}^{\infty} {dt\over t^{2}} t^{s}
\, \int_{\Sigma} {\rm Td}_{\Sigma}^{t} \wedge {\tr}_{\CR}\ e^{-t ( {\sigma} + {\psi} + F + {\tilde\bm})} \cr}}
where
\eqn\tdt{{\rm Td}_{\Sigma}^{t} = 1 + {t\over 2} c_{1}({\Sigma})}
and $\tilde\bm$ is a background twisted chiral superfield, associated to the flavor symmetry.  Somewhat confusingly, the term {\it twisted} applies to $\tilde\bm$ in two ways. First of all, in the physical supersymmetry, the flavor symmetry can be weakly gauged, producing a background vector multiplet ${\CV}_{\rm f}$. To this multiplet on associates the background twisted chiral multiplet ${\bf M}_{\rm f} = {\CD}_{+}{\bar\CD}_{-} {\CV}_{\rm f}$. Upon the topological twist the components of this superfield further change spin, producing the superfield $\tilde\bm$. We are interested only in the bosonic components of $\tilde\bm$, which are the scalar ${\tilde m}$ (the matrix of twisted masses), and the two form $S \cdot c_{1}({\Sigma})$, which is the matrix of spins $S$ of the formerly scalar components of the chiral matter multiplets, times the two-form of the worldsheet curvature, representing the first Chern class $c_{1}({\Sigma})$ of the tangent bundle to $\Sigma$. 

\subsubsec{The gauge contribution}

Analogously to the computation of ${\tilde W}_{\rm matter}$, ${\CW}_{\rm matter}$, ${\CU}_{\rm matter}$, the effects of the massive gauge multiplets, the $W$-bosons and their superpartners, can be interpreted using index theory. The idea is to consider the inclusion
\eqn\jinc{j: Bun_{T_{\IC}} \to Bun_{G_{\IC}}}
which is defined at the level of the moduli stacks of holomorphic principal bundles on a Riemann surface $\Sigma$. The moduli spaces of stable bundles, or unitary flat connections, usually considered by physicists, would not do here. One then applies the corresponding version of the Riemann-Roch-Grothendieck formula (here the contribution of Grothendieck is crucial). 
{}In physics language this corresponds to the idea, exploited in \BlThlgt\btverlinde\LosevCR, that one can choose (generically) a gauge ${\sigma}(x) \in {\bt}$, thereby reducing the gauge group locally from $G$ to $T$. In this way the non-abelian localization of \WittenTwoD\ becomes the usual localization with respect to the infinite dimensional gauge group of $T$-valued gauge transformations. 

At any rate, the calculation of the equivariant Euler class of the normal bundle to $Bun_{T_{\IC}}$ in $Bun_{G_{\IC}}$ reduces to the RRG formula: 
\eqn\nrml{Euler (T Bun_{G_{\IC}}/ T Bun_{T_{\IC}}) = Ch( H^{1} ({\Sigma}, {\bg}/{\bt} ))  = - \int_{\Sigma} Td_{\Sigma} {\rm Tr}_{{\bg}/{\bt}} e^{{\sigma}}} 
where we assumed the vanishing of the zero'th cohomology, $H^{0} = 0$.
From \nrml\ one computes:
\eqn\ggth{\eqalign{& \int_{\Sigma} {\CO}_{{\CW}_{\rm gauge}}^{(2)} + c_{1}({\Sigma})  {\CO}_{{\CU}_{\rm gauge}}^{(0)} = \cr
& \qquad
{d\over ds}\Biggr\vert_{s=0} {{\Lambda}_{\rm uv}^{s}\over {\Gamma}(s)} \int_{0}^{\infty} {dt\over t^{2}} t^{s}
\, \int_{\Sigma} {\rm Td}_{\Sigma}^{t} \wedge {\tr}_{{\bg}/{\bt}}\ e^{-t ( {\sigma} + {\psi} + F)} \cr}}
In the case of the three dimensional theory, we are interested not in the Euler class of the normal bundle, but in that of the exterior algebra: 
\eqn\nrmliii{\eqalign{& Euler \left({\bigwedge}_{y} \left( T Bun_{G_{\IC}}/ T Bun_{T_{\IC}}\right) \right)   = \cr
& {\exp} \sum_{n=1}^{\infty} {y^{n} \over n} Ch( {\psi}^{n} H^{1} ({\Sigma}, {\bg}/{\bt} ))  = {\exp}   \int_{\Sigma} Td_{\Sigma} {\rm Tr}_{{\bg}/{\bt}} {\rm log} \left( 1 -  y e^{{\sigma}} \right) \cr}}
The Kaluza-Klein resummed \ggth\ becomes:
\eqn\ggthiii{\eqalign{& \int_{\Sigma} {\CO}_{{\CW}_{\rm gauge}^{3d}}^{(2)} + c_{1}({\Sigma})  {\CO}_{{\CU}_{\rm gauge}^{3d}}^{(0)} = \cr
& \qquad
\sum_{n=1}^{\infty} {1\over n^2} \, \int_{\Sigma} {\rm Td}_{\Sigma}^{n} \wedge {\tr}_{{\bg}/{\bt}}\ e^{-2\pi {\ii} n  ( {\sigma} + {\psi} + F)} \cr}}
resulting in the Eqs. (3.33) and (3.35) below. 

\subsubsec{The cutoff dependence}

{}We need to say a few words about the ${\Lambda}_{\rm uv}$-dependence. Clearly, Riemann-Roch-Grothendieck formula does not have the ultraviolet cut-off in it. It is easy to compute the ${\Lambda}_{\rm uv}$-dependence of \asth:
\eqn\dimdiff{
{\rm log} {\Lambda}_{\rm uv} \left( {\rm dim}_{\IC}{\CM} - {\rm dim}_{\IC} Bun_{G_{\IC}} \right)
}
so we get a factor of $\Lambda_{\rm uv}$ per fermionic zero mode, the usual conversion between the topological and physical supersymmetric theory \nsvz.

\subsubsec{More examples of effective twisted superpotentials}

The formulas \effsp\wmatii\wmatiii\gsup\gsupiii\ give the examples of effective twisted superpotentials of the two and three dimensional theories. These formulas, as well as the formulas for the four dimensional ${\CN}=1$ theories compactified on the two-torus can be found in \NekrasovUI\NekrasovUH.

Another interesting example comes from the $\Omega$-deformed four dimensional ${\CN}=2$ theory, studied in \NekrasovRC. The four dimensional ${\CN}=2$ gauge theory can be deformed in the background with $U(1) \times U(1)$ isometry generated by the vector fields ${\Omega}_{1}, {\Omega}_{2}$, $[ {\Omega}_{1}, {\Omega}_{2}] = 0$, 
\eqn\omga{
{\Omega}_{a} = \left( x^{2a - 1} {\partial}_{x^{2a}} - x^{2a} {\partial}_{x^{2a-1}} \right), \qquad a=1,2}
by, roughly speaking, promoting the adjoint Higgs field to the differential operator:
\eqn\omgdf{{\sigma} \longrightarrow {\sigma} + {\epsilon}_{1} {\nabla}_{{\Omega}_{1}} + {\epsilon}_{2} {\nabla}_{{\Omega}_{2}}}
The ${\CN}=2$ super-Poincare super-algebra of flat Minkowski background is broken, by the $\Omega$-deformation, to the small superalgebra generated by a complex supercharge $\CQ$ with the commutation relation ${\CQ}^{2} =$ rotation generated by the complex vector field ${\epsilon}_{1}{\Omega}_{1} + {\epsilon}_{2}{\Omega}_{2}$. We shall use the gauge theory partition function, 
\eqn\zfun{{\CZ}({\ba}, {\epsilon}_{1}, {\epsilon}_{2}; {\bm}, {\bf\tau})}
i.e. the supersymmetric partition function of the theory on ${\IR}^{4}$ with appropriate boundary conditions at infinity, with the adjoint Higgs approaching the conjugacy class $\ba$ at infinity:
\eqn\sitoa{{\sigma} \left(  {r\to \infty} \right) \qquad \longrightarrow \qquad  {\ba} = {\rm diag}(a_{1}, \ldots , a_{N}) }
with the masses $\bm$ corresponding to the flavor symmetries
and the complexified gauge couplings $\tau$.  The $\CZ$-function contains the information, among other things, about the properties of the two dimensional effective theory obtained by the $\Omega$-deformation in two dimensions, i.e. with $( {\epsilon}_{1}, {\epsilon}_{2}) = ({\epsilon}, 0)$.

The effective twisted superpotential of the effective two dimensional theory is the sum of two terms: the universal part, which captures the ultraviolet gauge dynamics 
\eqn\wuniv{{\CW}^{\rm univ} ( {\sigma} = {\ba}, {\bm}, {\epsilon}) = {\rm Lim}_{{\epsilon}_{2} \to 0} \ {\epsilon}_{2} \, {\rm log} {\CZ}({\ba}, {\epsilon}_{1} = {\epsilon}, {\epsilon}_{2}; {\bm}, {\bf\tau}) }
 and the infrared one-loop part, which depends crucially on boundary conditions at infinity in the plane, transverse to the two-dimensional space-time \NekrasovXDA. Here the parameter $\epsilon$ of the $\Omega$-background is one of the twisted masses, it corresponds to the flavor symmetry of the two dimensional theory, which is the rotational symmetry of the four dimensional theory, corresponding to the rotations transverse to the two dimensional space-time. 
  
\subsubsec{Gravitational couplings}

Just like the effective twisted superpotential splits as a sum \effsp\ 
of the tree-level part, the one-loop contribution from the matter fields and the one-loop contribution from the massive $W$-bosons, so does the effective gravitational coupling
\eqn\effcu{{\CU}^{\rm eff} = {\CU}^{\rm uv} + {\CU}_{\rm matter} + {\CU}_{\rm gauge}}
Now, in \asth\ the scalar component of $\bm$ is equal to the twisted mass matrix $\tilde m$ while the two-form is equal to $c_{1}({\Sigma})$ times the matrix
$S$ of spins of the lowest components of the matter field after the twist (see \NekrasovUI\NekrasovUH\NekrasovRC). 
In two dimensions, i.e. for finite dimensional
$\CR$ we conclude:
\eqn\cueffm{{\CU}_{\rm matter} = {\tr}_{\CR}\ ( {\half} - S ) {\log} \left( { {\sigma} + {\tilde m} \over {\Lambda}_{\rm uv}} \right)} 
The massive gauge supermultiplets contribute:
\eqn\cueffg{{\CU}_{\rm gauge} = \sum_{{\alpha}} {\rm log} \left( { \langle {\alpha}, {\sigma} \rangle \over {\Lambda}_{\rm uv}} \right) = {\rm log}{{\Delta}^2 ({\sigma}) \over (-{\Lambda}_{\rm uv}^{2})^{{\rm dim}({\bg}/{\bt})/2}}}
the logarithm of the Vandermonde determinant, or the rational Weyl denominator. 

In three dimensions, the sum over the Kaluza-Klein
modes of matter fields gives (cf. \nrmliii):
\eqn\cueffiiim{{\CU}_{\rm matter}^{3d} = {\tr}_{{\CR}_{3}}
\ ( {\half} - S ) {\rm log}\left( 1 - e^{-2\pi R ({\sigma} +{\tilde m})} \right)}
The contribution of the massive vector multiplets in three dimensions sums up to (cf. \nrmliii):
 \eqn\cueffiiig{{\CU}_{\rm gauge}^{3d} = \sum_{\alpha} {\rm log}\left( 1 - e^{-2\pi R \langle {\alpha}, {\sigma} \rangle} \right)}
 which is essentially the Weyl denominator, or trigonometric Vandermonde determinant.

Observe that in two and three dimensions the matter contribution to the effective gravitational coupling ${\CU}_{\rm matter}$ is determined by the effective twisted superpotential:
\eqn\cufrcw{{\CU}_{\rm matter}({\sigma}; {\tilde m}, S) = \left( {\half} - S \right) \cdot {\partial\over {\partial\tilde m}} {\CW}({\sigma}; {\tilde m})} 
Let us also give the expression for the gravitational coupling of the two dimensional theory, obtained from the four dimensional ${\CN}=2$ theory by the two dimensional $\Omega$-deformation with parameter $\epsilon$:
\eqn\cuome{{\CU}( {\sigma} = {\ba}, {\bm}, {\epsilon}) = {\rm Coeff}_{{\epsilon}_{2}^{0}} \  \, {\rm log} {\CZ}({\ba}, {\epsilon}_{1} = {\epsilon}, {\epsilon}_{2}; {\bm}, {\bf\tau}) }
 at ${\epsilon}_{2} \to 0$. Here the relation like \cufrcw\ is spoiled by the non-perturbative effects of four dimensional instantons.

Now let us discuss the contribution of the massive vector fields. 
In two dimensions, 
\newsec{Topological partition function}
Once $\CW^{0}$ and ${\CU}^{0}$ are calculated, the generating function of correlators of $0$-observables of the chiral ring generators ${\CO}_{A}({\s})$ on the Riemann surface $\Sigma$ of genus $g$ is given by the sum over the generalized critical points of the  effective twisted superpotential:
\eqn\topart{Z_{g}(t)=\sum_{B} {\CH}({\sigma}_{B})^{g-1}\ e^{-\sum_{A} t_{A} {\CO}_{A}( {\sigma}_{B})}}
where  ${\sigma}_{B}$ are the solutions of the vacuum equations (Bethe equations)
\eqn\bethe{B: \qquad {1 \over {2 \pi \ii}}{{\p {\CW}^{0}}(\sigma) \over {\p \s}^{i}} = n_i \in {\IZ}} 
and ${\CH}({\sigma}_{B})$ is computed from $\CW$ and $\CU$ as follows:
\eqn\hsb{{\CH} ({\sigma}) = e^{-2 {\CU}^{0}({\sigma})}\cdot {\Det}_{ij}\, \Biggl\Vert{{\partial}^{2}{\CW}^{0} \over {\partial}{\sigma}_{i} {\partial}{\sigma}_{j}} \Biggr\Vert \cdot {\Delta}^{-2} ({\sigma}) }
and ${\Delta}({\sigma})$ is the Vandermonde determinant:
\eqn\des{{\Delta}({\sigma}) = \prod_{{\alpha}>0} \langle {\alpha} , {\sigma} \rangle.}
The second step is the analysis of the two dimensional Yang-Mills theory, which follows that of
  \WittenTwoD.
 
\subsec{Topological metric and the handle gluing operator}

Recall that the two dimensional topological field theory defines a commutative and associative algebra $\CA$. The generators of this algebra as of the vector space over $\IC$ are the linearly independent operators ${\CO}_{A}$
in the cohomology of the operator $\CQ$. They multiplication defines the structure constants:
\eqn\cabc{
{\CO}_{A} \cdot {\CO}_{B} = C_{AB}^{C} {\CO}_{C} + \{ {\CQ}, \ldots \} } 
Let us assume that the identity operator ${\CO}_{0} =1$ belongs to the algebra (this is an assumption of a version of compactness of the target space). 
The three-point function on the sphere defines a related $3$-tensor:
\eqn\thrf{\langle {\CO}_{A} (0) {\CO}_{B}(1) {\CO}_{C}({\infty}) \rangle_{S^{2}} = C_{ABC}}
Namely, let us define the {\it topological metric}
\eqn\etme{{\eta}_{AB} = C_{A B 0}, \ }
which we assume to be invertible, ${\eta}^{AB}{\eta}_{BC} = {\delta}^{A}_{C}$, 
then
\eqn\ctoe{C_{AB}^{C} = {\eta}^{CD}C_{ABD}}
The following element of the algebra is canonically defined:
\eqn\hgo{{\CH} = {\eta}^{AB} C^{C}_{AB} {\CO}_{C} = {\eta}^{AB} {\CO}_{A} \cdot {\CO}_{B} \in {\CA}}
It is called the {\it handle gluing operator}. Its significance is due to the following simple fact:
\eqn\zprt{Z_{g} = {\tr}_{\CA} \, {\CH}^{g-1}} 
The application of this formula to the three dimensional theory, obtained by deforming the minimally supersymmetric Yang-Mills theory by the Chern-Simons observable produces an alternative derivation of Verlinde formula, cf. \btverlinde\gerasimov\nikdiss\BaulieuNJ. Of course the physical derivation skips various subtleties, which are by now thoroughly addressed in \Faltings\Teleman. 
 
\subsec{Handle gluing operator in Landau-Ginzburg theory and in Gromov-Witten theory}
A prototypical example of the topological field theory is the Landau-Ginzburg theory of some chiral fields $X^{i}$, $i=1,\ldots , n$ with the superpotential $W(X)$ and the volume form 
$$
\Omega = {\Omega}(x) dx^{1}\wedge dx^{2} \wedge \ldots \wedge dx^{n} .  
$$
It is a B model,
with the chiral ring being the local ring of $W$:
\eqn\crlg{{\CA} = {\IC}[X]/\langle {\partial}_{1}W, \ldots , {\partial}_{n}W \rangle}
The three point function on the sphere given by:
\eqn\thrlg{\langle {\CO}_{f_{1}} {\CO}_{f_{2}} {\CO}_{f_{3}} \rangle = \oint_{\Gamma} {\Omega} \left( {{f_{1}(X) f_{2}(X) f_{3}(X)}\over {{\Omega}^{-1}(x){\partial}_{1}W \ldots {\partial}_{n}W}} \right)}
The handle gluing operator is given by, in this case (cf. \vafacr)
\eqn\hlg{{\CH} = {\Det} \left( {{\partial}^{2}W \over {\partial}X^{i} {\partial}X^{j}} \right) {\Omega}(x)^{-2}}

The Gromov-Witten theory is the A type topological sigma model with some target space $V$ coupled to topological gravity. The (twisted) chiral ring $\CA$ is the cohomology $H^{*}(V)$, with the deformed multiplication, called the quantum multiplication.  The topological metric
$\eta_{AB}$ is actually the classical intersection pairing. 
The three point function on the sphere given by:
\eqn\thram{\langle {\CO}_{{\omega}_{1}} {\CO}_{{\omega}_{2}} {\CO}_{{\omega}_{3}} \rangle = \sum_{{\beta} \in H_{2}(V, {\IZ})}
e^{-\int_{\beta} t} \int_{{\CM}_{0,3}(V, {\beta})} ev_{1}^{*}{\omega}_{1} \wedge ev_{2}^{*}{\omega}_{2} \wedge ev_{3}^{*}{\omega}_{3}}

\subsec{The handle gluing operator in the topological gauge theory}

We have effectively computed the operator $\CH$ in the topological gauge theories related to the Bethe/gauge correspondence  
 
\subsec{Implications for Gaudin conjecture}
  
There is an interesting basis in the twisted chiral ring coming from the Bethe/gauge correspondence:
\eqn\betheba{{\CO}_{A} \leftrightarrow {\Psi}(x; {\sigma})}
where ${\Psi}(x; {\sigma})$ is the wave function of the Bethe state $ |{\sigma}\rangle$ of the quantum integrable system \niklect. In other words this is the function on some space with the coordinates $x$, possibly discrete, solving the eigenvalue problem for the commuting quantum integrals of motion:
\eqn\commh{{\hat H}_{i,x} {\Psi}(x ; {\sigma}) = E_{i} ({\sigma}) {\Psi}(x ; {\sigma})}
where we indicated that the quantum Hamiltonians ${\hat H}_{i,x}$ act in the $x$-space. Their eigenvalues depend on the spectral parameters $\sigma$, which are the scalars in the vector multiplet
on the gauge side of the correspondence. Of course \commh\ does not fix the normalization of $\Psi (x; {\sigma})$. It is remarkable that there exists a natural normalization where the topological metric ${\eta}^{AB}$ acts as the PT-conjugation (Hermitian conjugation on the real slice):
\eqn\bethebaa{{\eta}^{AB}{\CO}_{B} \leftrightarrow {\Psi}^{*}(x; {\sigma})}
so that the handle gluing operator becomes:
\eqn\hgo{{\CH} = {\CH}({\sigma}) = \sum_{x} {\Psi}^{*}(x; {\sigma}) {\Psi}(x; {\sigma})}
The formula \hsb\ then becomes the statement of Gaudin's conjecture on the norm of the Bethe states, proven by V.~Korepin in many interesting examples \ref\korone{V.E. Korepin,
``Calculations of norms of Bethe wave-functions'',
Comm. Math. Phys.
86
(1982), 391.} (see also e.g. \ref\kortwo{V. E. Korepin and O. I. P\^atu, http://insti.physics.sunysb.edu/~korepin/montreal.pdf.}). 
\bigskip
\noindent {\bf Acknowledgments.}  We thank A. Gerasimov, S. Gukov and T. Hausel for discussions. We especially thank S. Gukov for reminding us the importance of having the general formula for the handle gluing operator published.  Such a formula has been reported in various lectures \niklect\ , but in full generality it never appeared in the previous papers except for some cases studied in \higgs, \gerasimovshatashvili, \gstwo. 

SSh thanks A. Alekseev for the hospitality at the Section Mathematiques of the University of Geneva during April-May 2014.

Research of NN was supported in part by RFBR grants 12-02-00594, 12-01-00525, by Agence Nationale de Recherche via the grant
ANR 12 BS05 003 02,  by Simons Foundation, and by Stony Brook Foundation.

The work of SSh is supported in part by the Science Foundation Ireland under the RFP program, by the ESF Research Networking Programme ``Low-Dimensional
Topology and Geometry with Mathematical Physics (ITGP)'' and by the grant number 141329 of the Swiss National Science Foundation.

\footatend\vfill\supereject\immediate\closeout\rfile\writestoppt
\baselineskip=14pt\centerline{{\bf References}}\bigskip{\frenchspacing%
\parindent=20pt\escapechar=` \input refs.tmp\vfill\eject}\nonfrenchspacing
\bye